\documentclass[11pt,twoside]{article}
\usepackage{asp2010}

\resetcounters

\begin{document}

\title{VOGCLUSTERS: an example of DAME web application}

\author{Marco Castellani, $^1$ Massimo Brescia, $^2$, Ettore Mancini $^3$, 
Luca Pellecchia $^3$ Giuseppe Longo $^3$ }
  \affil{$^1$INAF - Astronomical Observatory of Rome,\\
  $^2$INAF - Astronomical Observatory of Capodimonte (Naples) \\
  $^3$Dept. of Physics, University Federico II (Naples)}

\begin{abstract} We present the alpha release of the 
VOGCLUSTERS web application,
specialized for data and text mining on globular clusters. 
It is one of the web2.0 technology based services of Data Mining \&
Exploration (DAME) Program, devoted to mine and explore heterogeneous information
related to globular clusters data.

\end{abstract}

\section{Introduction}

Along with the progress of the instruments for astronomical investigation of our Galaxy, 
the amount of data available for each globular clusters (GC) in the Milky Way 
is also growing at a steady rate. 
In this very moment, several instruments from earth and from space are 
restlessly working to improve our detailed knowledge of such objects, 
which are of utterly importance for our understanding of the 
status and evolution of our Galaxy.
Moreover, since some of the older objects in the Universe are contained 
in globular clusters, they have a key role also in many cosmological topics.

The importance of globular clusters 
for a broad range of astronomical studies 
has been already addressed by Harris (1996). 
In that seminal paper (that introduced the famous online 
parameter compilations for galactic globular clusters) 
he also stressed the importance of having readily available up-to-date 
list of parameters for these unique objects. 
After all these years, his words appear even truer: 
not only we have new and more reliable parameters for a great part of the known clusters, 
but - thanks also to modern surveys conducted in bands different from the visible, such as 2MASS 
(Skrutskie et al. 2006) - several other objects keep going to increase our list of Milky Way 
clusters (e.g., Froebrich et al. 2008, Moni Bidin et al. 2011).

However, at the present day, such data appears still scattered 
among the various papers and/or are reported in different 
(and for the most part not-homogeneous) 
web pages and online catalogues. Such resources typically cannot be easily put in connection by 
people who want to make researches involving all kind of data.

Nowadays it appears more and more important to delineate a strategy in order 
to present the relevant information on a given cluster, or a range of clusters, 
in a single source and under a well-defined standard. 
This represents a necessary step to disclose a wide range of new investigations: 
the more we find ways to explore the parameters space, 
the more we'll be able to make useful science 
investigating upon the correlations of those set of data.

In the following it is presented a brief introduction to a project born to meet these 
requirements, named VOGCLUSTERS \footnote{http://dame.dsf.unina.it/vogclusters.html}. 

\section{A first step: the galactic globular clusters database}

The Galactic Globular Clusters Database (gclusters)
is an online resource \footnote{http://gclusters.altervista.org}, focused on presenting, 
in an organized way, a comprehensive list of bibliography, 
parameters and data for each of the known GCs 
of the Milky Way (Castellani 2007). 

From the technical point of view, {\it gcluster} 
it is a "classical" dynamical web site, and can be considered as a prototype of VOGCLUSTERS. Initially built around the Harris’ compilation, 
it is designed to allow a more flexible fruition of available data, 
allowing a number of operations on data, such as ordering 
clusters according to the value of a given parameter, select objects whose parameters 
falling in a given interval, display related bibliography and colour magnitude diagrams, or 
even drop a note pertinent to that cluster, to be displayed online for other user. 

Data are collected from several different sources, such as NASA Astrophysics Data System (ADS), Digital Sky Survey, 
related papers and websites, etc. 

\section{A further step: moving to VOGCLUSTERS web application}

The goal of the project VOGCLUSTERS is the development of a web application 
specialized in data and text mining activities for astronomical archives 
related to galactic and extragalactic GCs. 
Main services are employed for the simple and quick 
navigation in the archives and their 
manipulation to correlate and integrate internal scientific information. The archives are uniformed
under Virtual Observatory standard and constraints, in order to provide an homogenous and 
flexible environment, virtually capable of interactions with an ever growing amount of
external resources.
At variance with {\it gcluster},
the project has not to be intended as a straightforward website, 
but as a {\it web application}. 

A website usually refers to the front-end interface through which the public interact with your 
information online. Websites are typically informational in nature with a limited amount of 
advanced functionality. Simple websites consist primarily of static content where the data 
displayed is the same for every visitor and content changes are infrequent; more advanced 
websites may also have management and interactive content. A web application, or equivalently 
{\it Rich Internet Application (RIA) }
usually includes a website component but features additional 
advanced functionality to replace or enhance existing processes. The interface design objective 
behind a web application is to simulate the intuitive, immediate interaction a user experiences 
with a desktop application.

\section{The DAME framework}

DAME (DAta Mining \& Exploration) is an international collaboration for data mining and machine
learning research program, by exploiting the “web2.0” technologies and exposing a series of
applications and services for e-science communities. In particular, since the beginning,
it is specialized in astrophysical services.
Its products are hence basically focused on data/text mining on massive data sets with machine
learning methods, on top of an hybrid distributed computing infrastructure (Brescia et al. 2010).
The DAME design architecture is implemented following the standard LAR 
(Layered Application Architecture) strategy, which leads to a software system based 
on a layered logical structure, where different layers communicate with each other 
via simple and well-defined rules:

\begin{enumerate}
\item Data Access Layer (DAL): the persistent data management layer, 
responsible of the data archiving system, including consistency and reliability maintenance.
\item Business Logic Layer (BLL): the core of the system, responsible of the management 
of all services and applications implemented in the infrastructure, including information 
flow control and supervision.
\item User Interface (UI): responsible of the interaction mechanisms between the BLL and the users, 
including data and command I/O and views rendering.
\end{enumerate}

\section{Details and status of the VOGCLUSTERS web app}

The VOGCLUSTERS web application is hosting an integrated specialized toolset in the DAME
infrastructure,
then taking fully advantage of its features.
In particular, its integration deals with the technological solutions adopted, 
derived from DAME Web Application Suite strategy and requirements. 
For example, concerning the user access security policy, VOGCLUSTERS shares 
the DAME user archive and, being on top of pre-existing CLOUD-GRID hybrid architecture, 
it inherits integrity and security levels. Inside DAME, all VOGCLUSTERS authorized 
users are protected against privacy and data consistency violations.

\begin{figure}
\epsscale{1.3}
\plotone{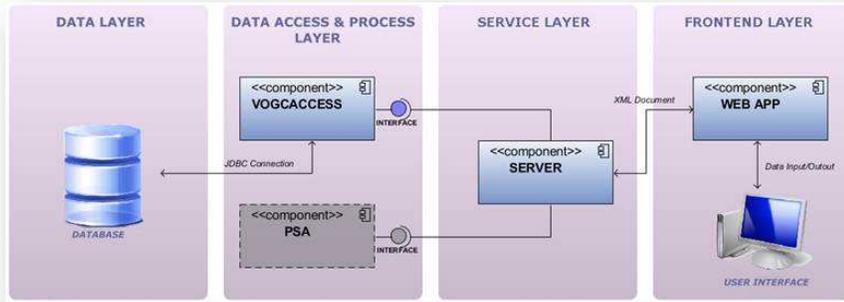}
\caption{The VOGCLUSTERS architecture: each box shows a different layer of the application}
\label{disknopert}
\end{figure}

The VOGCLUSTERS application is at an advanced phase of development. Its main features are:

\begin{enumerate}

\item fully adoption of the DAME layered model;

\item developed adopting the Google Web Toolkit (GWT). 
GWT make possible to use java language (which is rapidly reaching the status of a  
standard  for a great number  of scientific applications and projects, so that it is 
widely diffused among researchers), 
having java classes translated into an AJAX powered website. 
AJAX is an acronym for Asynchronous JavaScript and XML; it represents 
a group of interrelated web development methods used on the client-side 
to create interactive web applications. This choices allow us to develop powerful features
while maintaining low the requirements of time and human resources.
\item As usual for a DAME application, it is developed according to the standards of 
Virtual Observatory (VO). In this way, present and future archives in VOGCLUSTERS can take 
advantage of a well-defined 
environment expressly ideated for the easy exchange of information between different 
resources. 

\end{enumerate}

We consider that the knowledge of a GC is made of observational parameters, 
sometimes including also an history of the research 
evolution on their values.
The user has the possibility 
to visualize and navigate the GCs archives (galactic and extra-galactic objects) and make 
correlation, in an constantly increasing number of ways, asking also for on-the-fly plots of
different combination and/or selection of data (coming soon).
The information are transparently presented to the user in a simple and attractive
web interface (by merging worldwide distributed VO databases and registries). 
So far the user has the possibility to integrate/update these 
parameters with own values and reports (new values, comments, images, diagrams, references etc.).
Results obtained by a user will not be locked in the application: soon it will be possible 
to save them in a number of formats, like FITS, EPS, PDF, etc.

The web application foresees three different categories of users, 
respectively, the administrators, qualified (registered) and 
generic (not registered) users. The main differences between these user categories 
are related to the information manipulation rights. In particular the web application 
can be passively navigated by all users (generic category), can be integrated/updated 
by registered users (already authenticated within DAME Suite) in terms of own data management 
and can be re-engineered by administrators (typically DAME working group members). 
The first stable beta release of the application should be available by
the end of next October.


\section{References}
\bibliography{aspauthor}
\noindent
Brescia, M. et al.2010 {\it in press} (arXiv:1010.4843v2) \\
Castellani, M. 2007, MemSait 79,2 \\
Froebrich, D. et al. 2008, MNRAS, 390, 1598 \\
Harris, W.E. 1996, AJ, 112, 1487 \\
Moni-Bidin, C. et al. {\it in press} (arXiv:1109.1854) \\
Skrutskie, M.F. et al. 1996, AJ, 131, 1163
\end{document}